\documentclass[12pt]{JHEP3}

\pdfoutput=1
\usepackage{bm}
\usepackage{epsfig}
\usepackage{citesort}
\usepackage{graphicx}

\def\be{\begin{equation}}
  \def\ee{\end{equation}}
\def\bea{\begin{eqnarray}}
  \def\eea{\end{eqnarray}}

\title{Assumptions of the primordial spectrum and cosmological parameter estimation} 

\author{Arman Shafieloo\\
	Department of Physics, University of Oxford\\ 
	1 Keble Road, Oxford, OX1 3NP, UK\\
	\emph{and} \\
	Institute for the Early Universe, Ewha Womans University\\ 
	Seoul, 120-750, South Korea\\
	E-mail: \email{arman@ewha.ac.kr}}

\author{Tarun Souradeep\\
	Inter University Centre for Astronomy and Astrophysics\\ 
	Post Bag 4, Ganeshkhind, Pune, 411007, India\\
	E-mail: \email{tarun@iucaa.ernet.in}}

     
\abstract{The observables of the perturbed universe, CMB anisotropy and large
structures, depend on a set of cosmological parameters, as well as,
the assumed nature of primordial perturbations. In particular, the
shape of the primordial power spectrum (PPS) is, at best, a well
motivated assumption. It is known that the assumed functional form of
the PPS in cosmological parameter estimation can affect the best fit
parameters and their relative confidence limits.  In this paper, we
demonstrate that a specific assumed form actually drives the best fit
parameters into distinct basins of likelihood in the space of
cosmological parameters where the likelihood resists improvement via
modifications to the PPS.  The regions where considerably better
likelihoods are obtained allowing free form PPS lie outside these
basins. In the absence of a preferred model of inflation, this raises
a concern that current cosmological parameters estimates are strongly
prejudiced by the assumed form of PPS.  Our results strongly motivate
approaches toward simultaneous estimation of the cosmological
parameters and the shape of the primordial spectrum from upcoming
cosmological data. It is equally important for theorists to keep an open mind towards early universe scenarios that produce features in the PPS.}

\begin{document}

\section{Introduction}                        
\label{sec:introduction}
Precision measurements of anisotropy and polarization in the Cosmic
Microwave Background (CMB), in conjunction with observations of the
large scale structure, suggest that the primordial density
perturbation is dominantly adiabatic and has a nearly scale invariant
spectrum~\cite{sel04,sper_wmap06_dunkly_08}. This is in good agreement with most
simple inflationary scenarios which predict nearly power law or scale
invariant forms of the primordial
perturbation~\cite{inflation1,inflation2,inflation3}.  However,
despite the strong theoretical appeal and simplicity of a featureless
primordial spectrum, our results highlight that the determination of
the shape of the primordial power spectrum directly from observations
with minimal theoretical bias would be a critical requirement in
cosmology.

The observables of the perturbed universe, such as, CMB anisotropy
galaxy surveys and weak lensing etc., all depend on a set of
cosmological parameters describing the current universe, as well as,
the parameters characterizing the presumed nature of the initial
perturbations.  While certain characteristics of the initial
perturbations, such as, the adiabatic nature, tensor contribution,
can, and are, being tested independently, the shape of the primordial
power spectrum remains, at best, a well motivated assumption.

It is important to distinguish between the cosmological parameters
that describe the present universe, from that characterizing the
initial conditions, specifically, the primordial power spectrum (PPS), $P(k)$. However, it is
prevalent in cosmological parameter estimation to treat the two sets
identically. Based on sampling on a coarse grid in the cosmological
parameter space, we have already shown that the CMB data is sensitive
to the PPS~\cite{prd08}.  The best fit cosmological parameters with
free form PPS have much enhanced likelihoods and the preferred regions
significantly separated from the best fit parameters obtained with
assumed power-law PPS.  It is also known that specific features in the
PPS can dramatically improve the fit to data (eg. ref.~\cite{rjain08},
and references therein).


In this paper, we bring forth another important issue introduced by
our prior ignorance about the PPS. While, known correlations between
cosmological parameters is always folded into parameter estimation,
the analogous situation for $P(k)$ is not as widely appreciated.  An
assumed functional form for the PPS, is equivalent to an analysis with
a free form PPS, where, say, $P(k)$, is estimated in separate $k$
bins, but, then one imposes strong correlation between the power in
different bins. As we show in this work, the assumed form
(equivalently, the implied correlations in $P(k)$ at different $k$,)
drive the significant number of degrees of freedom available in the
cosmological parameters to adjust into suitable specific
combinations. Hence, the assumed form of PPS could be dominant in
selecting the best fit regions (eg., ref.~\cite{hunt_sarkar} where
features in the PPS lead to very different best fit cosmological
parameters). For specific functional forms, the corresponding best-fit
models lie entrenched in distinct basins in the parameter space.  Our
results show that in these basins, the likelihood is remarkably robust
to variations in the PPS. We conclude that there are sufficient degrees
of freedom in the cosmological parameters to mould the fit around the
constraints imposed by the assumed form of the PPS.


In this paper, we elucidate this issue in the context of CMB data from
WMAP for three different well known assumed forms of the primordial
spectrum, $P(k)$, (i.) scale-invariant Harrison-Zeldovich (HZ) with $P(k)=A_s$, (ii.)
scale-free Power-law (PL) with $P(k)=A_s[\frac{k}{k_*}]^{n_s-1}$ and, (iii.)  Power law with running(RN) with $P(k)=A_s[\frac{k}{k_*}]^{n_s(k_*)-1+\frac{1}{2}ln(k/k_*)dn_s/dlnk}$ where $k_*$ is a pivot point.
The methodology and analysis is described in $\S$\ref{meth}. The
results are given in $\S$\ref{res} and we give the conclusion of
our work in $\S$\ref{concl}.



\section{Method and Analysis}                        
\label{meth}
The angular power spectrum of CMB anisotropy, $C_l$, is a convolution
of the PPS, $P(k)$ generated in the early universe with a radiative
transport kernel, $G(l,k)$, determined by the current values of the
cosmological parameters.  The precision measurements of $C_l$, and the
concordance of cosmological parameters measured from other
cosmological observations allow the possibility of direct recovery of
$P(k)$ from the observations. In our analysis we use an improved
(error-sensitive) Richardson-Lucy (RL) method of deconvolution to
reconstruct the optimized primordial power spectrum at each point in
the parameter
space~\cite{rich72lucy74,baug_efs9394,prd04,prd07,prd08}.  The RL
based method has been demonstrated to be an effective method to
recover $P(k)$ from $C_l$ measurements~\cite{prd04,prd07,prd08} (look at ref.~\cite{max_zal02_bridle03_pia03_bump05_kog05_leach05_bridges08} for some other reconstruction methods).

In this paper, we study the improvement in likelihood allowed by an
`optimal' free-form PPS at points in the cosmological parameter space
around the best-fit region for the three different assumed form of
PPS, viz., Harrison-Zeldovich (HZ), power-law (PL) and power-law with
running (RN).  We apply our deconvolution method to reconstruct an
`optimal' form of the PPS at each point~\cite{prd08}. 

Markov-chain Monte-Carlo (MCMC) samples of parameters provide a fair
sampling of the parameter space around the best-fit point. We use the
MCMC chains generated based on 3 year data by the WMAP team for parameter estimation with
HZ, PL and RN forms of the PPS. We reconstruct the optimized PPS for
each point of these chains and obtain the `optimal' PPS likelihood
based on the reconstructed spectrum.

We limit our attention to the flat $\Lambda$CDM cosmological model and
consider the four dimensional parameter space, $\Omega_b h^2$, $\Omega_{0m} h^2$, $h$ and $\tau$. This corresponds to a minimalistic
"Vanilla Model", a flat $\Lambda$CDM parametrized by six parameters 
($n_s$, $A_s$, $H_0$, $\tau$, $\Omega_{b}$, $\Omega_{dm}$).
In case of Harrison-Zeldovich (HZ) PPS assumption, $n_s= 1$, leaving
only $5$ parameters. The case of assuming a constant running in the
spectral index (RN), $n_{run}$, leads to $7$ parameters.  The
dimensionality in the three models is different solely due to the
parameters of the assumed PPS. Hence, in our analysis we always have a
four-dimensional space of cosmological parameters (since we recover
the optimal PPS).

In order to represent the likelihood in a four dimensional parameter
space we find it convenient to define a normalized distance, $\rho$,
between two points, 
\begin{equation}
\rho(a,b)= \sqrt{\Sigma_i
{(P_i^a-P_i^b)^2}/{(\sigma_i^b)^2}},
\label{rhodef} 
\end{equation}

where
$P_i^a$ and $P_i^b$ are the value of $i^{\rm th}$ cosmological
parameter at point `a' and point `b', respectively. To ensure that
equal separations along different parameters have a similar meaning,
we divide $P_i^a-P_i^b$ by standard deviation $\sigma_i^b$ at point
`b'.  We assign point `b' to be a best fit point where $\sigma_i^b$
are the $1\sigma$ confidence limits derived by WMAP team from the
corresponding MCMC chains. Since we are primarily interested in
studying the region around the best-fit point, $\rho$ provides a
convenient definition of distances to other points with respect to
it. (Note, the `distance' $\rho$ is `asymmetric' in `a' and `b' when
$\sigma_i^b \neq \sigma_i^a$ and should interpreted accordingly).
\section{Results}                        
\label{res}

The simplest characterization of the likelihood landscape, ${\cal
L}(P_i)$, around the best-fit point is to study its behavior as $\rho$
increases with separation from the best-fit point.
The trend in the likelihood can then be compared for two cases --
assuming a form of primordial spectrum, or allowing a optimal free
form. (We 
use the effective chi-square,
$\chi^2\equiv -2\ln {\cal L}$ instead of ${\cal L}$.)

\FIGURE{
\includegraphics[width=.66\textwidth,angle=0]{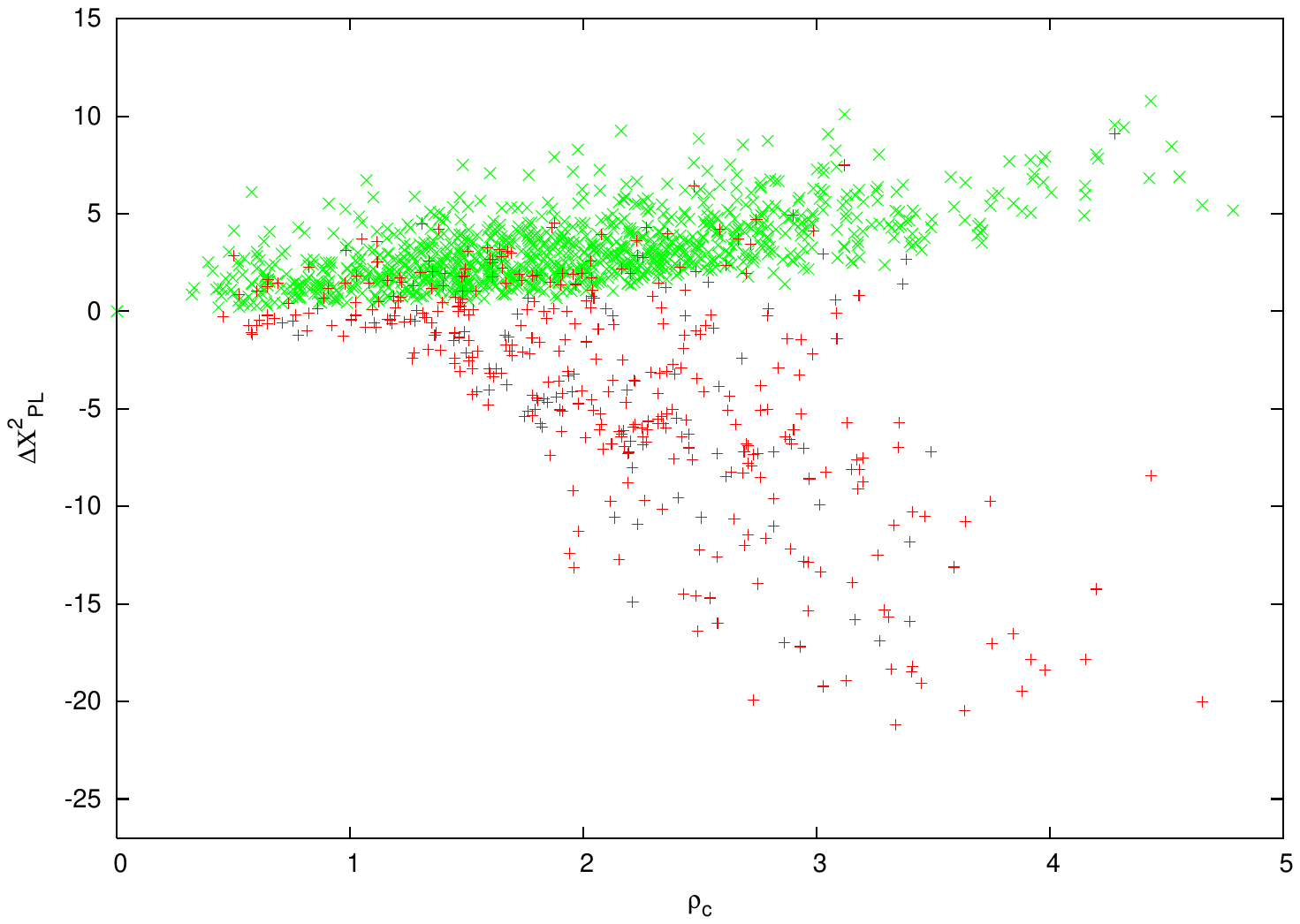} 
\includegraphics[width=.66\textwidth,angle=0]{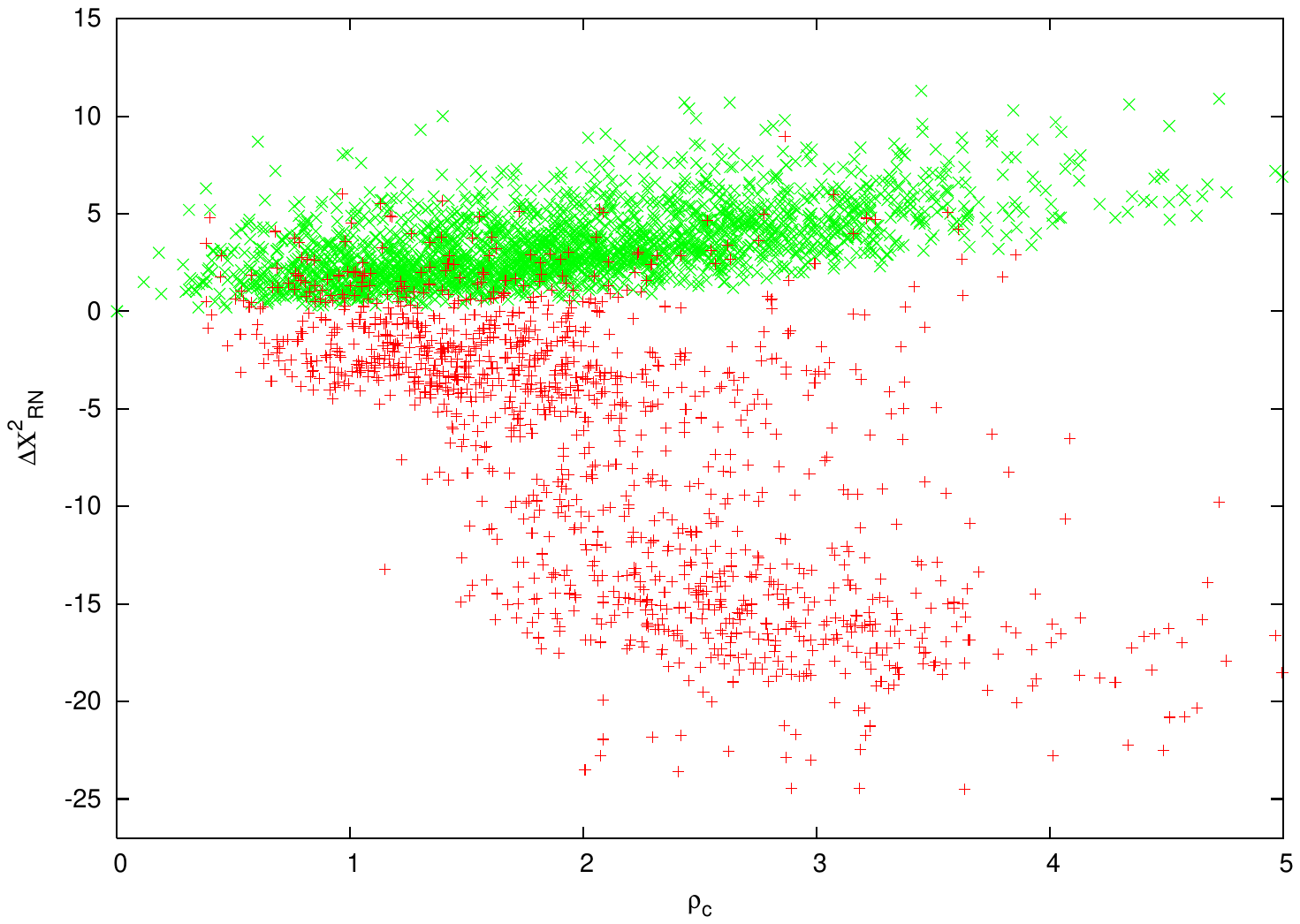} 
\caption{The panels show the comparative
scatter-plots of relative $\chi^2$, with, and without, optimal $P(k)$,
versus normalized distance, $\rho_c$, (eqn.~\ref{rhodef}) in parameter
space of the sample points for sub-samples of the MCMC chains
generated by the WMAP team.  Green crosses show the $\Delta \chi^2$
relative to the best fit value. Red pluses mark the same points in the
parameter space but with $\chi^2$ derived after `optimization' of the
primordial power spectrum.  The {\em top} and {\em bottom} panels
corresponds to MCMC chains assuming `Power Law' (PL) form and `running
power law'(RN) forms PPS, respectively. The obvious absence of red
points with significantly negative $\Delta\chi^2$ for $\rho_c < 1$
mark the basins for each assumed PPS where, no (or, minor) improvement
in likelihood is seen even invoking a free-form `optimal' PPS. The
basins for three assumed PPS are non-overlapping. For comparison, in
the PL case (Upper panel), the distances to the best-fit HZ model
$\rho_{(HZ,PL)} = 6.89$ and RN Model $\rho_{(RN,PL)} = 4.85$,
respectively.  In the RN case (Lower panel), distance to best-fit HZ
and PL models are $\rho_{(HZ,RN)} = 11.09$ and $\rho_{(PL,RN)}=2.44$,
respectively. \label{fig_PL_RN}}}

We now define $\rho_c$ as the distance between each point 
in the given MCMC samples to the best-fit point.
For each point, we compute the effective $\chi^2$ difference, $\Delta
\chi^2$ (i.e., twice the relative log-likelihood) with respect to this
best fit point, both, for the likelihood obtained under the assumed PPS,
and, with a free form PPS (the optimal PPS recover in our
deconvolution).
Fig.~1 
shows scatter-plots of $\Delta \chi^2$ vs
$\rho_c$ for the case of power-law (PL) and running power-law (RN)
assumptions of the primordial spectrum. Green crosses show the
expected behavior that locally the likelihoods worsens with $\rho_c$
as points depart from the best-fit parameters.  In the other hand, the
red pluses mark the same points in the cosmological parameter space,
but for the $\Delta \chi^2$ obtained under free-form optimal PPS.
It is clear from Fig.~1,
that a free-form optimal PPS
can very markedly improve the likelihood relative to that in assumed
form PPS. What is more remarkable is that the improvement through
optimal free-form PPS is suppressed in a basin around $\rho_c <
1$. This is apparent in the absence of red plus marks near the lower
left corners of the plots.

It is also interesting to mention that the basins for the three
assumed forms of the PPS are very distinct and non-over-lapping. The
parameter distances between the best fit points assuming HZ, power law
and power-law with running forms of the primordial spectrum are quite
large. We have $\rho_{(PL,HZ)} = 14.56$, $\rho_{(RN, HZ)} = 43.79$,
$\rho_{(HZ,PL)} = 6.89$, $\rho_{(RN,PL)} = 4.85$, $\rho_{(HZ,RN)} =
11.09$ and $\rho_{(PL,RN)} = 2.44$.  
It is important also to note that the best fit point obtained under
one assumed form of PPS may be disfavored with a high confidence by
another assumption.

\begin{figure*}[!t]
\centering
\begin{center}
\vspace{-0.05in}
\centerline{\mbox{\hspace{0.in} \hspace{2.1in}  \hspace{2.1in} }}
$\begin{array}{@{\hspace{-0.3in}}c@{\hspace{0.3in}}c@{\hspace{0.3in}}c}
\multicolumn{1}{l}{\mbox{}} &
\multicolumn{1}{l}{\mbox{}} \\ [-0.5cm]
 \includegraphics[scale=0.3, angle=0]{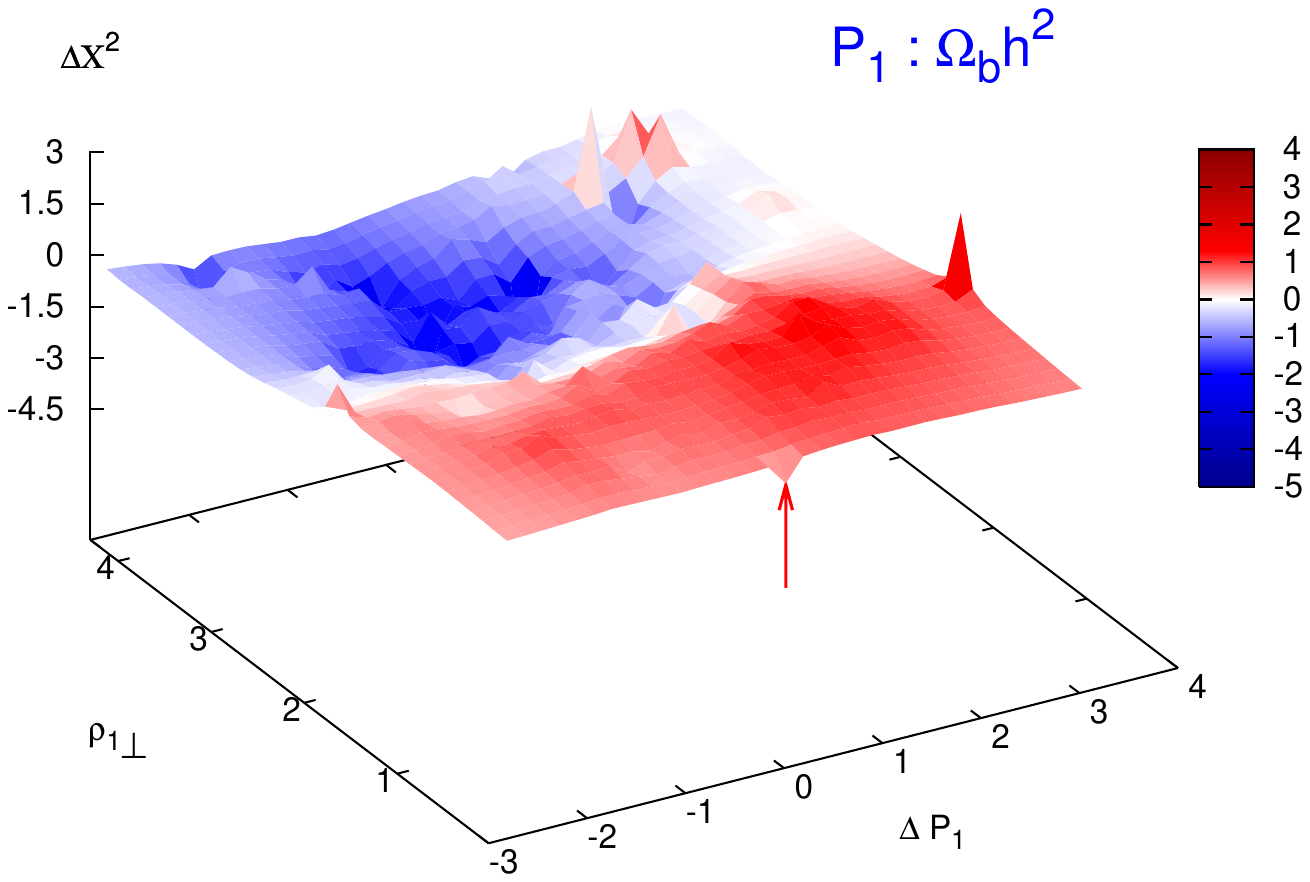}
 \includegraphics[scale=0.3, angle=0]{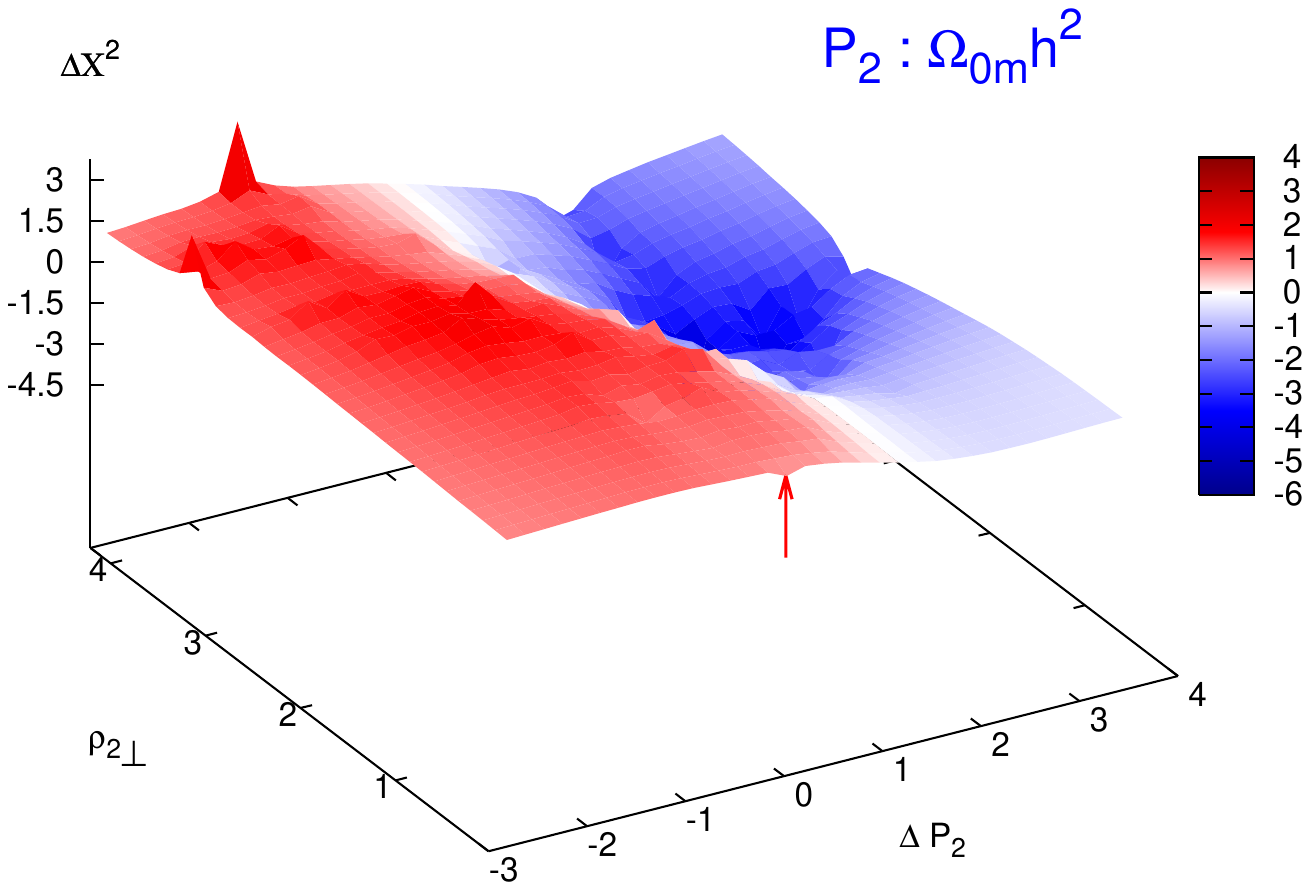}
\end{array}$
$\begin{array}{@{\hspace{-0.3in}}c@{\hspace{0.3in}}c@{\hspace{0.3in}}c}
\multicolumn{1}{l}{\mbox{}} &
\multicolumn{1}{l}{\mbox{}} \\ [-0.5cm]
\includegraphics[scale=0.3, angle=0]{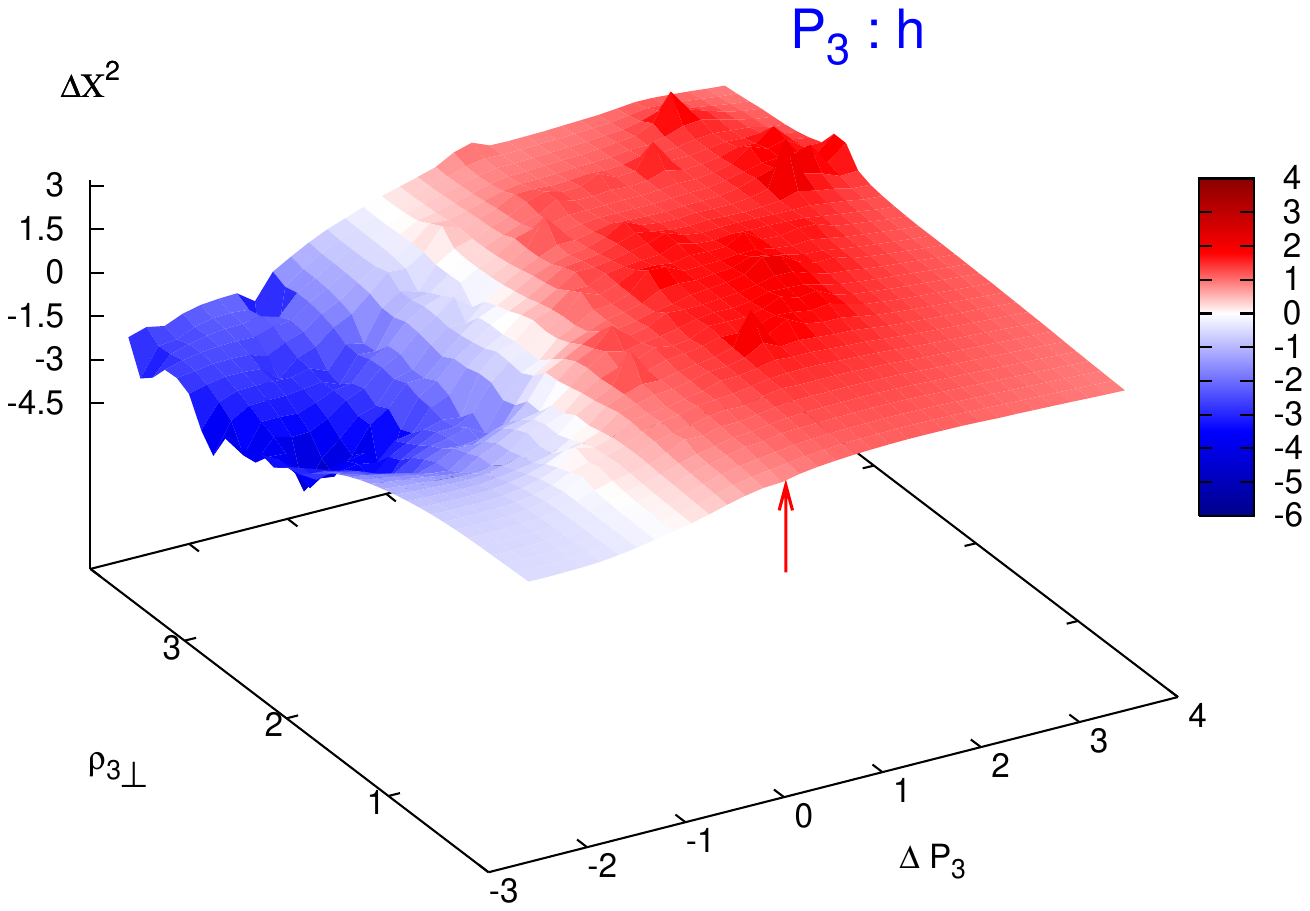}
  \includegraphics[scale=0.3, angle=0]{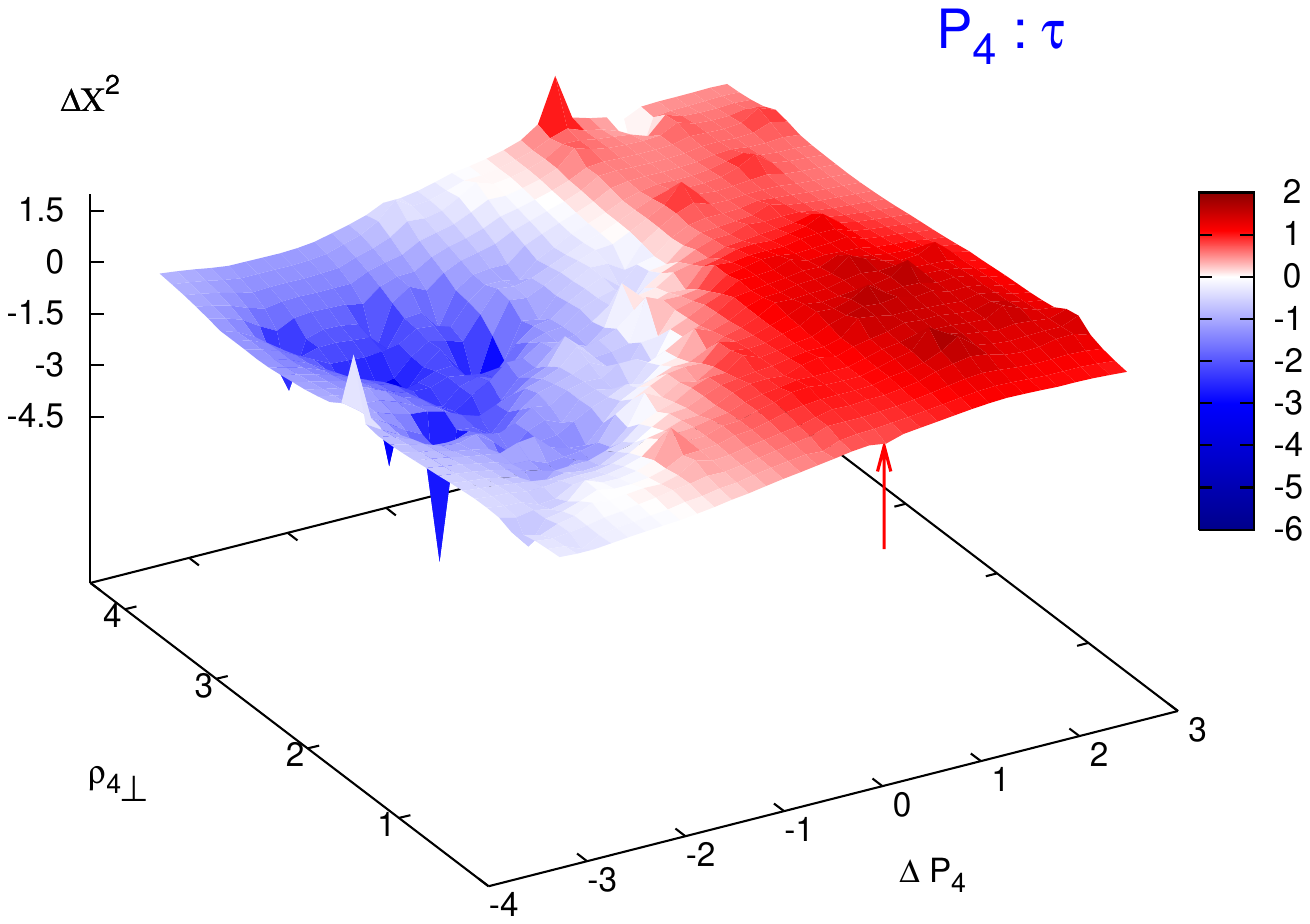}
\end{array}$
\end{center}
\caption{\small A $2D$ surface representation of the optimized $\Delta
\chi^{2}$ around the best-fit point for Power law PPS case for the
four parameters.  For each parameter, $i$, $\Delta P_i$ measures
separation along the parameter, and $\rho_{i_\perp}$ measures the
separation in the three other parameters.  Regions in the parameter
space with $\Delta \chi^2 > 0$ are shown by red color and are
separated by a white band (representing $\Delta \chi^2 \approx 0$)
from the regions with $\Delta \chi^2 < 0$ shown by blue color. Red
plateaus represent the regions where allowing the free form primordial
spectrum does not improve the likelihood. Red arrows show the position
of best fit point assuming PL form of PPS.} \label{fig_3D}
\end{figure*}

As mentioned above, in the basins around the best-fit points it is
very difficult to get a significantly better likelihood allowing for a
free form PPS.
It is instructive to explore the nature of these basins and the trends
of likelihood assuming the free form PPS for each of parameters
$\Omega_b h^2$, $\Omega_{0m} h^2$, $h,$ and $\tau$. To do so, for each
parameter, $i$ we split the separation, $\rho$, between points, $a$
and $b$ in the parameters into a separation $\Delta P_i =
{(P_i^a-P_i^b)}/{\sigma_i^b}$, along the parameter and the
`perpendicular' distance $\rho_{i_\perp}= \sqrt {\sum _{j\ne i}
{(P_j^a-P_j^b)^2}/{(\sigma_j^b)^2}}$ measuring the separation in the
other three parameters.

Fig.~2
shows for the PL PPS case, a $2D$ surface representation of the
optimized $\Delta \chi^{2}$ around the best fit point plotted against
$\Delta P_i$ and $\rho_{i_\perp}$ for each of the parameters. We have
weighted the neighbouring sample points by their Euclidean distance in
the parameter space to assign an average likelihood at each point.
The color palette is chosen such that red (blue) regions have poorer
(better) likelihood than the reference value of best-fit model. The
white regions have likelihood comparable to the best fit value.  In
this representation, the figures clearly show that in all cases, there
is a plateau in the parameter space (the red regions) enclosing the
best fit point where a free form PPS does not improve the
likelihood. The location of the best-fit points 
are marked by red arrows. Outside these basins, there are blue regions
where optimal free form PPS leads to very significant improvement in
the likelihood.  (However, note that these are far from being the
global minima for optimal PPS cosmological parameter estimates -- as
shown 
in ref.~\cite{prd08}, there are models with much higher likelihood.)

The plots in Fig.2,
also supplement the PL plot in Fig.~1 (top),
by indicating the direction in the
parameter space from the best fit models where likelihood resists
improvement against modifications to the PPS.

We should note that correlation between the cosmological parameters are visible looking at our results. The likelihood surfaces in Fig. 2 show at which points in the parameter space we may have the similar likelihood that is somehow representing the correlation between the parameters. The 4 cosmological parameters used in our analysis are the
basic parameters in CMB analysis and though there might be some correlations
between them (in derivation of likelihood) they are fundamentally
independent parameters. However, this has not been our main concern in this paper so we have not gone much in to the details. Our main point is to show how we may end up to another corners of the parameter space by assuming different forms (or a free form) of the primordial spectrum. We should also clarify that the choice of $\rho$ used
in this paper is somehow arbitrary in a sense that we want to have an
estimation of the distance between different points in the parameter
space. One could use another form of $\rho$ but the results could not be
significantly different. 

Another important point to mention is about the degrees of freedom in our likelihood analysis. One should note that it is not justifiable to define the degrees of freedom for our non-parametric reconstructed free form of the primordial spectrum. This
issue has been discussed earlier in~\cite{prd04,prd07,prd08}. It has been shown though that one cannot derive a very good likelihood at any point in the parameter space having this freedom of choosing the form of primordial spectrum. This fact is the base of this work which we try to optimize the form of the primordial spectrum to derive the
best likelihood at each point. One should realize that this freedom is for
all points in the parameter space so if we cannot improve the likelihood
in parts of the parameter space even though allowing the free form of the
primordial spectrum, this shows  the strong setting of the other
parameters (cosmological parameters). Our analysis is clearly a non Bayesian analysis and we show that setting the strong priors on the form of the primordial spectrum can results to ambiguities in derivation of the cosmological parameters and can be potentially misleading.

\section{Conclusion}                        
\label{concl}

 In this paper we have shown that the assumed form of
the primordial power spectrum (PPS) plays a key role in the
determination of cosmological parameters.  In fact, the functional
form of the PPS forces the best-fit cosmological parameters to
specific preferred basins of high likelihood to the data.  These
estimated cosmological parameters are then significantly biased. It is
similar to a case where in an $N$ dimensional parameter space of a
model, we fix the values of $m$ parameters ($m<N$) and vary the other
$N-m$ parameters to fit an observation. The resultant best fit values
of these $N-m$ parameters can be very different, depending on the
values assigned to the fixed $m$ parameters. If there is no good reason
to select a particular set of fixed values, the determination of the
rest of the parameters remains under question. Assumption of assumed (say,
power-law) form of the primordial spectrum can also be interpreted as
a very strong, specific correlation between $P(k)$ at different
$k$. This assumption is similar to setting values for the $m$
parameters with specific correlations. We surmise that the assumed form of
the PPS could be the dominant reason that in the basins for each
assumed form it was not possible to achieve a marked improvement in
$\chi^2$ by allowing optimal free-form PPS (see Fig.2). It is very important to note that despite allowing a free-form for the primordial spectrum, not all cosmological models (i.e., all points in the parameter space) can be fitted equally well to the data . We clearly show that some points in the cosmological parameter space fit the WMAP CMB data significantly better than the other points, by `optimizing' the likelihood over a free form of the primordial spectrum. We conjecture that the positive definiteness of the primordial spectrum does not allow us to fit all the points in the parameter space to the data equally well, and some points will have a better fit to the data. Hence, the result that we do not get a good likelihood for some points in the parameter space has nothing to do with being trapped in a local minima. We do not employ any global minimization (or, sampling) algorithm/technique where this could be be an issue. 
We should also clarify that our analysis in this paper, has not meant to be a
consistency check of the standard power-law form of the primordial spectrum (or
any form of P(k)). It is known that power-law form of the primordial spectrum is
well consistent with the data~\cite{verde_peiris09,bridges09,hamann_shafieloo09}. Our main point here in this paper is to emphasize the fact that in the process of cosmological parameter estimation, assumptions regarding the functional form of the primordial spectrum can be extremely critical. Not only does preferentially select best fit values from distinct basins in the parameter space, the apparent `robustness' to variations in $P(k)$ within these basins may be misleading.



In summary, we show that the apparently `robust' determination of
cosmological parameters under an assumed form of $P(k)$ may be
misleading and could well largely reflect the inherent correlations in
the power at different $k$ implied by the assumed form of the PPS.  We
conclude that is very important to allow for deviations from scale
invariant, scale free or, simple phenomenological extensions of the
same, in the PPS while estimating cosmological parameters. This
provides strong motivation to pursue approaches that simultaneously
determine both, the cosmological parameters, as well as, the
primordial power spectrum from observations. The rapid improvement in
cosmological observations, such as, the CMB polarization spectra, holds
much promise towards this goal. It is not unlikely that early universe scenarios that produce features in PPS could in fact be favored by data.


\acknowledgments{We have used WMAP data and the likelihood code provided by WMAP team in the Legacy Archive for Microwave Background Data Analysis (LAMBDA)
website~\cite{LAMBDA}.  Support for LAMBDA is provided by the NASA
Office of Space Science. In our method of reconstruction we have used
a modified version of CMBFAST~\cite{cmbfast}. We acknowledge use of
HPC facilities at IUCAA. A.S acknowledge the support of the European Research and Training Network MRTPN-CT-2006~035863-1~(UniverseNet) and Korea World Class University grant no. R32-10130.}

\appendix


\begin{thebibliography}{99}

\bibitem{sel04} U. Seljak. et al.,  Phys. Rev. {\bf D 71}, 103515 (2005).
\bibitem{sper_wmap06_dunkly_08} D. Spergel et al., Astrophys.J.Suppl. 170, 377 (2007); J. Dunkley et al., Astrophys.J.Suppl, {\it in press} (arXiv:0803.0586)
\bibitem{inflation1} A. A. Starobinsky, Phys. Lett, {\bf 117B}, 175   (1982).
\bibitem{inflation2} A. H. Guth \& S.-Y. Pi, Phys. Rev. Lett., {\bf 49}, 1110  (1982).
\bibitem{inflation3} J. M. Bardeen, P. J. Steinhardt \& M. S. Turner,
Phys. Rev. {\bf D 28}, 679 (1983).
\bibitem{prd08} A. Shafieloo \& T. Souradeep, Phys. Rev. {\bf D 78},
023511 (2008); T. Souradeep \& A. Shafieloo, Prog. of
Theor. Phys. Suppl. {\bf 172}, 156 ,(2008).
\bibitem{rjain08} R. Jain et al., JCAP, {\it in press} (arXiv:0809.3915)
\bibitem{hunt_sarkar} P. Hunt \& S. Sarkar, Phys. Rev. {\bf D 70}, 103518 (2004); {\it ibid}, {\bf D 76}, 123504 (2007).
\bibitem{rich72lucy74}B. H. Richardson, J. Opt. Soc. Am., {\bf 62}, 55
(1972); L. B. Lucy, Astron. J., {\bf 79}, 6 (1974).  .
\bibitem{baug_efs9394} C. M. Baugh and G. Efstathiou, MNRAS {\bf 265},
145 (1993); {\it ibid}, {\bf 267}, 323 (1994).
\bibitem{prd04} A. Shafieloo and T. Souradeep, Phys Rev. {\bf D 70}, 043523 (2004).
\bibitem{prd07} A. Shafieloo et al. Phys. Rev. {\bf D 75}, 123502 (2007).
\bibitem{max_zal02_bridle03_pia03_bump05_kog05_leach05_bridges08} M. Tegmark and M. Zaldarriaga, Phys. Rev. {\bf D66}, 103508, (2002); S. L. Bridle et. al., MNRAS. {\bf 342} L72 (2003); P. Mukherjee and Y. Wang, Astrophys.J. {\bf 599}  1 (2003); D. Tocchini-Valentini, Y. Hoffman, J. Silk, MNRAS. {\bf 367} T1095 (2006); N. Kogo, M. Sasaki, J. Yokoyama, Prog.Theor.Phys. 114, 555, (2005); S. Leach, MNRAS. {\bf 372} 646 (2006); M. Bridges et. al., (arXiv:0812.3541)
 
\bibitem{verde_peiris09}
H. Peiris and L. Verde, Phys. Rev. {\bf D 81}, 021302 (2010).
\bibitem{bridges09}
M. Bridges et al. MNRAS. {\bf 400} 1075B (2009).

\bibitem{hamann_shafieloo09}
J. Hamann, A. Shafieloo and T. Souradeep, JCAP {\bf 1004}, 010 (2010). 


\bibitem{LAMBDA} Legacy Archive for Microwave Background Data Analysis [http://lambda.gsfc.nasa.gov/].
\bibitem{cmbfast} U. Seljak \&  M. Zaldarriaga, Astrophys.J. 469, 437 (1996).

\end{thebibliography}
\end{document}